\documentclass{kapproc} % Computer Modern font calls

\usepackage{t1enc}
\usepackage{procps} 
\usepackage{amssymb}

\usepackage[dvips]{graphicx}

\let\footnote\savefootnote

\upperandlowercase
\kluwerbib % will produce this kind of bibliography entry:

\begin{document}

\articletitle{Comparing Dynamical and Stellar Pop-\\ulation Mass-to-Light
  Ratio Estimates}

%\Articlesubtitle{This is an Article Subtitle}

\author{Roelof de Jong}
\affil{Space Telescope Science Institute, Baltimore, U.S.A.}
\email{dejong@stsci.edu}

\author{Eric Bell}
\affil{Max-Planck-Institut f\"ur Astronomie, Heidelberg, Germany}
\email{bell@mpia.de}

\begin{abstract}
We investigate the mass-to-light ratios of stellar populations as
predicted by stellar population synthesis codes and compare those to
dynamical/gravitational measurements. In Bell \& de Jong (2001) we
showed that population synthesis models predict a tight relation
between the color and mass-to-light ratio of a stellar population. The
normalization of this relation depends critically on the shape of the
stellar IMF at the low-mass end. These faint stars contribute
significantly to the mass, but insignificantly to the luminosity and
color of a stellar system. In Bell \& de Jong (2001) we used rotation
curves to normalize the relation, but rotation curves provide only an
upper limit to the stellar masses in a system. Here we compare stellar
and dynamical masses for a range of stellar systems in order to
constrain the mass normalization of stellar population models. We find
that the normalization of Bell \& de Jong (2001) should be lowered by
about 0.05-0.1 dex in {\em M/L}. This is consistent with a Kroupa (2001),
Chabrier (2003) or a Kennicutt (1983) IMF, but does not leave much
room for other unseen components.

\end{abstract}


\section{Introduction}
In Bell \& de Jong (2001) we showed that stellar population models
predict a strong correlation between an optical color of a stellar
population and its mass-to-light ({\em M/L}) ratio (see also e.g., Bell et
al.~2003; Portinari et al.~2004). We showed that the {\it slope} of this
relation is rather insensitive to the exact details of the star
formation history and chemical enrichment of the stellar population
(except for recent star bursts) and to dust reddening, owing to 
the well-known age/metallicity/dust degeneracy. 
Furthermore, the color--{\em M/L} slope
is also rather insensitive to the IMF used. Yet, the 
{\it normalization} of the color--{\em M/L} relation is highly
IMF dependent, shifting up and down depending on how many stars are 
present at the low-mass end of the stellar IMF (these stars contribute
significantly to the mass of a population, but insignificantly 
to its luminosity and color).

In Bell \& de Jong (2001) we used maximum disk rotation curves to
constrain the normalization of the color--{\em M/L} relation. The predicted
stellar population masses derived from the color--{\em M/L} relation should
never over-predict the observed dynamical masses derived from rotation
curves. However, while rotating gas in a disk galaxy is a very simple
dynamical system and hence a clean constraint, rotation curves of disk
galaxies have the disadvantage that they only provide an upper limit
to {\em M/L} ratios once we accept that dark matter may be present in disk
galaxies. There is no guarantee that there is no unseen matter
contributing to the dynamical mass within the radius where the maximum
disk is constrained, be it baryonic (e.g., cold molecular gas) or
non-baryonic. Hence, rotation curves only provide an upper limit to
the normalization of the color--{\em M/L} relation. Here we compare
dynamical masses and masses predicted by stellar population modeling
of a variety of stellar systems in order to constrain the
normalization of the color--{\em M/L} relation\footnote{In principle, any
of these systems could have a dark component co-spatial with the
stellar light (in some cases this is rather unlikely), 
and hence all comparisons are
strictly speaking upper limits to the relation.}.

\section{Comparing dynamical and population {\em M/L} estimates}

In order to compare dynamical and stellar population masses we have to
make a number of assumptions:
\begin{itemize}
\item The IMFs of the stellar populations in the different objects are
  the same, notwithstanding the large range in object scale sizes and masses
  involved. 
\item The stellar population models used are accurate in a relative
  sense (not necessary in absolute calibration).
\item The stellar systems in question have not 
  selectively lost (or accreted) stars in a
  particular mass range. 
\item Where necessary we use the HST Key Project distance scale.
\end{itemize}

We will now go through a number of dynamical/gravitational versus
stellar population mass comparisons, and express the range of allowed
population {\em M/L} ratios in terms of the IMF normalization used in Bell
\& de Jong (2001), i.e.\ a Salpeter x=1.35 IMF between 0.1 and 125
$M_\odot$ reduced in mass by a factor 0.7\footnote{We do not explicitly
include a mass contribution for objects with masses less than 0.1
$M_\odot$; as argued later, the contribution from brown 
dwarf or planetary regime objects to the stellar {\em M/L} is expected
to be 0.04\,dex or less. }.

{\em Globular Clusters} At first sight globular clusters seem ideal
targets to compare dynamical and stellar population masses: their
stellar populations, dust corrections and dynamics are simple, and
they are unlikely to contain large amounts of dark matter near their
centers. However, mass segregation has resulted in centers of clusters
being dominated by massive stars, the outer parts by lower mass
stars. The outer stars are subsequently more likely to be stripped by
interaction with the galaxy, making it even harder to get a good
sampling of the original full IMF. Detailed dynamical modeling of
Galactic globular clusters shows clear evidence of these effects, with
{\em M/L} changing with radius (e.g., Gebhardt \& Fischer 1995).

When we compute the dynamical core {\em M/L} ratios of Galactic globular
clusters following McLaughlin (2000) and compare those to single burst
%{\em P\'EGASE} 
{\em PEGASE} 
(Fioc \& Rocca-Volmerange 1997) models using the colors
and metallicities of Harris (1996), we find that the dynamical {\em M/L}
values are much lower (by about 0.23 dex) than predicted by the single
burst models of a 10--12\,Gyr old population. Alternatively, we can
follow a more simplified approach by using virial masses (Pryor \&
Meylan 1993), which are more representative of the total globular
cluster. We find that the 12\,Gyr stellar population masses are lower
by 0.10 dex than the virial masses when using our Bell \& de Jong
(2001) IMF normalization, albeit with a large scatter of 0.20 dex rms
(comparable to the uncertainties in the dynamical {\em M/L} values).

In recent years it has also become possible to measure virial masses
of globular clusters in other nearby galaxies. This has the advantage
that it is easier to get integrated properties the globular clusters
and many objects at the same distance, but as disadvantage the limited
accuracy that can be reached, even with 8\,m class telescopes. The
results of extra-galactic globular clusters are still inconclusive,
with dynamical masses of Cen A as measured by Martini \& Ho (2004) being
more massive than our stellar population model predictions by 0.08
dex, but the dynamical masses of M33 (Larsen et al.~2002) being 0.27
dex lower than predicted.

{\em Elliptical galaxies} Recently, Cappellari et al.~(2006) have
performed a detailed analysis of dynamical and stellar populations
masses of a sample of early-type galaxies.  Their integral field
spectrograph SAURON data allows them to derive accurate dynamical
masses using Schwarzschild modeling and stellar population masses
using line-strength indices modeling. Using a Kroupa (2001) IMF and
Vazdekis et al.~(1999) stellar population synthesis models, they
find that old, fast rotating elliptical galaxies have dynamical and
stellar population masses that are very similar. Younger, fast
rotating elliptical galaxies have smaller stellar population masses
than dynamical masses, but they can be made to agree by assuming that
the young ages are the result of a superposition of a dominant, old
massive population and a small, young population. However, slowly rotating,
old massive elliptical galaxies seem to have higher dynamical than stellar
population {\em M/L} ratios, a discrepancy that cannot be solved by a
super-position of young and old populations, because the population
already is old according to the line indices. 

Cappellari et al.~(2006) argue that under the assumption that the IMF
is the same for all galaxies this must mean that these massive, slowly rotating
galaxies have a significant dark matter within their effective radius
where the dynamical measure was made. However, once we accept that some
elliptical galaxies must have a dynamically significant amount of dark
matter in their central region, we cannot exclude that all elliptical
galaxies have dark matter contributing to their central
dynamics. Therefore, the comparison of stellar population and
dynamical masses in elliptical galaxies becomes an upper limit to the
normalization of the ``IMF mass'', identical to the maximum disk
rotation curve constraint. In terms of this mass normalization, 
this argues for a $\sim 0.05$\,dex lower normalization that used by 
Bell \& de Jong (2001), given that the
Vazdekis models include masses down to 0.01 $M_\odot$.

{\em Maximum disk rotation curves:} As described above, we used
maximum disk rotation curve limits to normalize the color--{\em M/L}
relation in Bell \& de Jong (2001). In Kassin, de Jong \& Weiner
(2006) we have repeated this analysis, but we expanded the Verheijen
(1997) Ursa Major cluster sample with 34 luminous galaxies and
improved the treatment of shifting the mass models to another
distance. We compared the maximum disk values to the updated
color--{\em M/L} relations of Bell et al.~(2003) and find that the Bell \&
de Jong (2001) normalization is fully consistent with this expanded
data set. The normalization may at best be 0.05 dex higher to account
for the scatter in the color--{\em M/L} relation.

{\em Minimum disk rotation curves:}
While most galaxy rotation curves are fairly smooth, some show enough
structure to allow determination of a lower limit to a stellar {\em M/L}
to explain these structures under the assumption that the dark matter
component is smooth (e.g., Noordermeer et al.~2004). Such analysis is
complicated by the unknown intrinsic distribution of dark matter, the
effect of adiabatic contraction, and rotation curve uncertainties (including 
non-circular motions). 
Using NGC\,157 (Kassin et al.~2006) we find a
lower limit of -0.3 dex with respect to the Bell \& de Jong (2001)
normalization to explain the strongly declining rotation curve of this
galaxy. However, the large asymmetries and hence large errorbars on
the rotation curve of this galaxy limits the usefulness of this
galaxy. More suitable systems (mainly early-type spiral galaxies with
falling rotation curves) are studied by Noordermeer (2006).

{\em Disk velocity dispersions:} The mass and scale height distribution
determine the vertical velocity dispersion of a self-gravitating
disk. Thus, to determine a disk mass from a measured vertical velocity
dispersion in a face-on system we have to make assumptions about the
(unobservable) vertical stellar distribution, while for edge-on
systems, where we can measure the vertical stellar distribution, we
have to relate the observed radial and tangential velocity dispersion
to the (unobservable) vertical velocity dispersion (Bottema
1997). Recently, Kregel et al.~(2005) used velocity dispersions of a
sample of 15 edge-on galaxies to determine a dynamical mass
Tully-Fisher relation and compared it to the stellar population mass
Tully-Fisher relation of Bell \& de Jong (2001). They find an offset
of about -0.24 dex, assuming a vertical-to-radial velocity dispersion
ratio ($\sigma_z/\sigma_R$) of 0.6. However, the determined {\em M/L}
ratio scales quadratically with the poorly known $\sigma_z/\sigma_R$
ratio.  To the best of our knowledge, only 3 measurements of 
$\sigma_z/\sigma_R$ have been made to date, ranging between 
0.5 and 0.9 
(Gerssen, Kuijken, \& Merrifield~2000).  Without substantially better 
understanding of the behavior of $\sigma_z/\sigma_R$ as a function 
of galaxy properties,
it is unclear that one can place competitive constraints on stellar
{\em M/L} ratios using this method.
%The difference is also reduced
%when bulge masses are included in the comparison.

\begin{figure}
\includegraphics[width=6cm,viewport=23 48 532 390]{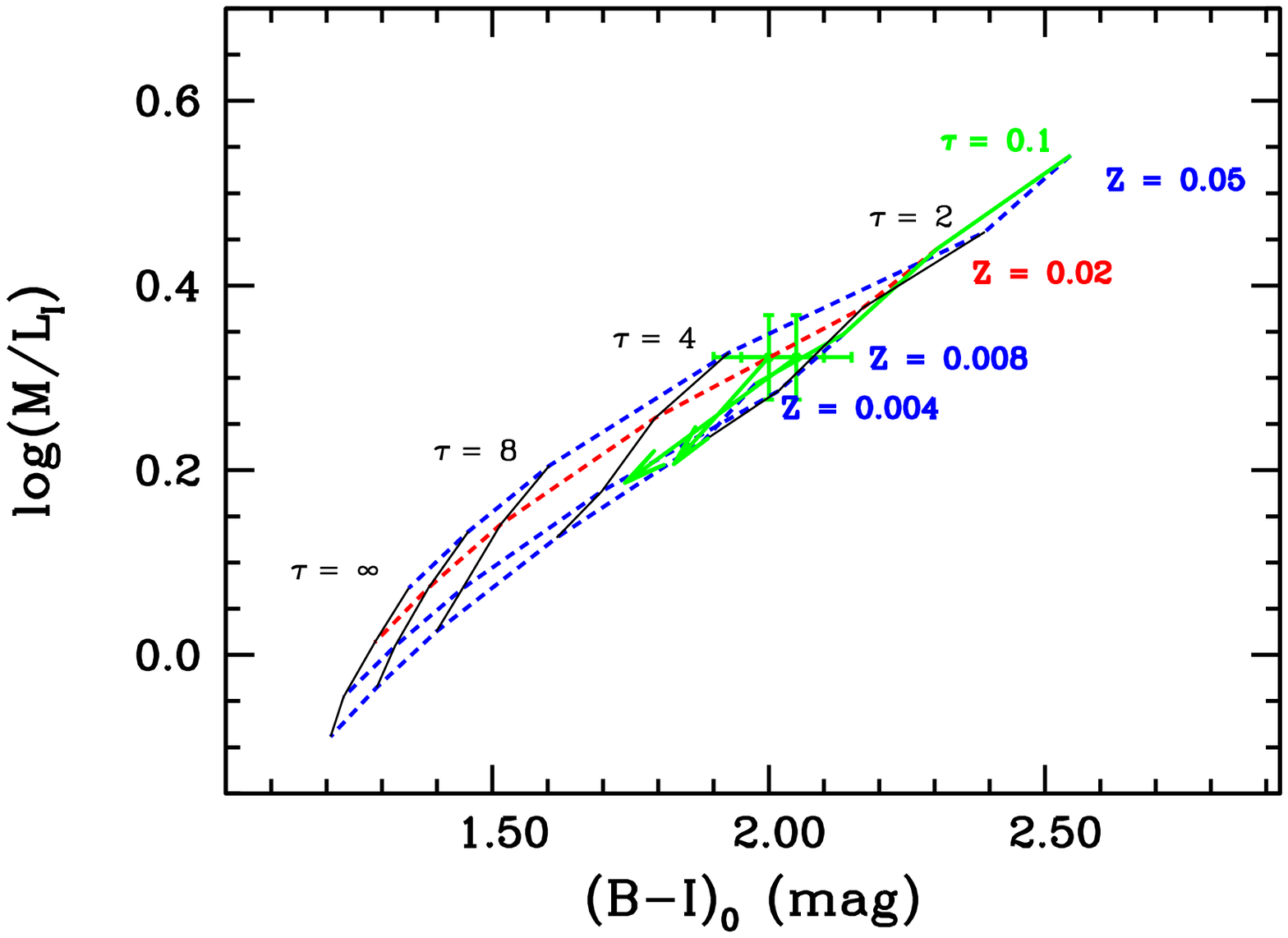}
\includegraphics[width=6cm,viewport=23 48 532 390]{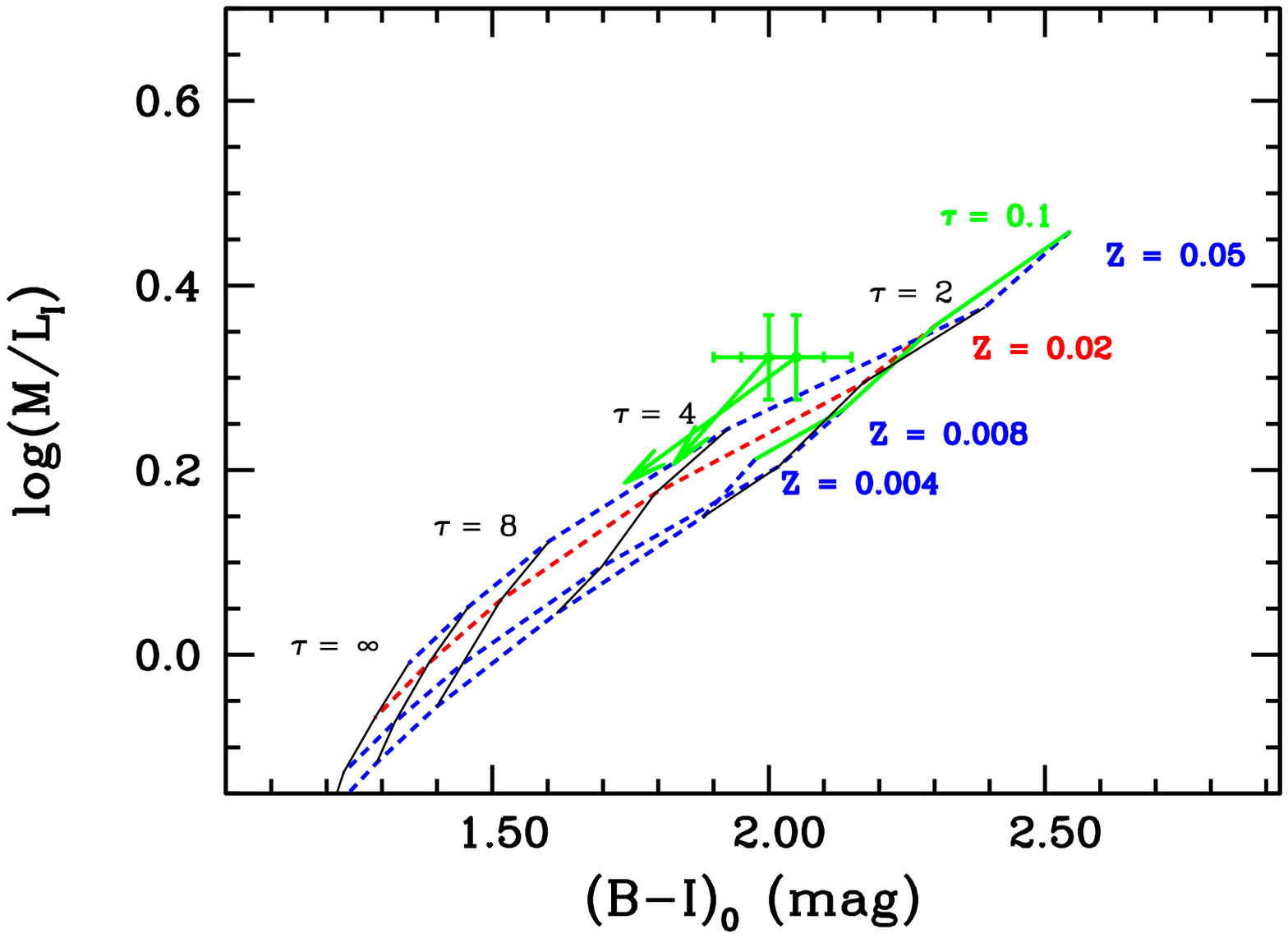}
\caption{Comparing the Weiner et al.~(2001; 2004) bar dynamics {\em M/L}
  constraints to P\' egase stellar population models (Fioc \&
  Rocca-Volmerange 1997). We plot 12 Gyr old exponential decaying star
  formation rate models with different decay rates $\tau$ connected by
  solid lines, different metallicities by dotted lines, as
  indicated. The two data points for NGC\,4123 and NGC\,3095 are
  indicated with formal errorbars. The arrows on the data points
  indicate the reddening for these galaxies according to Tully et
  al.~(1998). On the left we show the Bell \& de Jong (2001)
  normalization, at the right a 0.1 dex lower normalization.
\label{M_Lbar}
}
\end{figure}

{\em Bar streaming motions:} A galactic bar moving through the interstellar
medium creates streaming motions and often a shock, the size of which
depends somewhat on the pattern speed of the bar, but mostly on the
mass of the bar. Weiner et al.~(2001; 2004) obtained $H\alpha$ velocity
fields of NGC\,4123 and NGC\,3095 and modeled these with fluid-dynamical
models. Their models only permit a limited range in
stellar {\em M/L}, such that the galaxies are close to maximum disk. In
Fig.\,\ref{M_Lbar} we compare the local bar colors and the derived
{\em M/L} values of these two galaxies to the stellar population models of
a range in metallicity and with exponentially decaying star formation
rates. We show the models normalized at the Bell \& de Jong (2001)
value on the left, reduced by 0.1 dex in {\em M/L} on the right. The
models cover a limited area in these diagrams, showing the
age-metallicity degeneracy that makes the color--{\em M/L} relation work in
the first place.

We expect the central region to suffer from extinction, and we have
plotted indicative dereddening vectors on the measured data
points. These vectors were derived from Tully et al.~(1998) global
galaxy reddening values, and the extinction in the central region may
be somewhat higher. The left panel, where the Bell \& de Jong
normalization is used, shows that the raw and reddening-corrected
stellar {\em M/L} ratios are consistent with the model normalization.
The right-hand panel, with the model stellar {\em M/L} values
decreased by 0.1 dex compared to Bell \& de Jong (2001), is
just consistent with the reddening-corrected data. 
The bar streaming motions modeling therefore provides
some of the strongest constraints on our color--{\em M/L}
normalization, allowing only a range of $\sim$0.2 dex.

P\'erez et al.~(2004) confirm the analysis of Weiner et al.~(2001) to
the extent that, for the two out of their sample of five galaxies for
which they could derive {\em M/L} constraints, the barred galaxies had to
be close to maximum disk (at least 80\% stellar mass contribution in
bar region).

{\em Spiral arm streaming motions:} In a similar fashion, streaming
motions can be used to estimate the mass in a spiral density
wave. Clearly this is a more challenging exercise, as arm-induced
shocks and streaming motions are much weaker than those
induced by bars.  Kranz et al.~(2003) 
studied five high surface brightness
galaxies with long-slit spectra and optical/near-IR surface
photometry.  They can only weakly constrain stellar {\em M/L} (their Table 3), 
and find {\it i)} most of their sample have maximum disk {\em M/L} values
consistent with the Bell \& de Jong calibration, and {\it ii)}
most high $v_{\rm rot}$ disks are consistent with close to maximum disk, 
whereas the lower rotation velocity disks could be 
substantially sub-maximal (with 
less than $\sim$60\% of the disk mass coming from stars
within 2.2 disk scalelengths).

{\em Strong galaxy lensing:} Strong gravitational lensing provides an
estimate of the gravitational mass within the lens area with rather
straightforward modeling. To compare the gravitational mass of the
lens with its stellar population mass we have to correct the observed
colors using k-corrections or better yet, redshift the model spectra
to the lens redshift and calculate a new color--{\em M/L} grid in the
observed bands. Furthermore, we have to realize that galaxies are
younger at higher redshift and the color--{\em M/L} relation will
shift. For simplicity we can model this with exponentially declining
star formation rate models (which becomes an increasingly poorer
approximation at higher redshift, because starbursts will become
relatively more important). Such exponentially declining models ---
started 12\,Gyr ago --- have color--{\em M/L} relations in {\it restframe
  colors} that are decreased by 0.15 dex at redshift 1 compared to the
$z$=0 relation. 

To minimize these corrections a lensing galaxy at low redshift should
be used in the gravitational to stellar mass comparison. Furthermore,
the lensed images should be of small angular separation, surrounding
only the central region of the lens galaxy, which is most likely to be
dominated by stellar mass. We should keep in mind that any masses
derived from lensing, like maximum rotation curves, only provides an
upper limit to the stellar population mass estimates, as significant
dark matter may be present in the centers of some galaxies as also
suggested by the Cappellari et al.~(2006) elliptical galaxies result.

Indications from the first studies to satisfy these criteria are
encouraging.  Smith et al.~(2005) present an example of a low redshift
($z$=0.0345), tight lens, finding a $M/L_I$$\sim$1.8,
$M/L_B$$\sim$4.7.  Using the reported F475W and F814W observed
magnitudes, galactic foreground corrections, and k-corrections
assuming a non-evolving ancient galaxy template (solar metallicity and
$\sim$12\,Gyr old), we find excellent agreement with the predicted
stellar {\em M/L} values, as usual accounting for gas recycling from ageing
stellar populations ({\em M/L}$_{I,{\rm pred}}$$\sim$1.8, {\em M/L}$_{B,{\rm
    pred}}$$\sim$4.5).  Koopmans et al.~(2006) analyze an extensive
sample of 15 lenses with z=0.06--0.33.  In this case, the
dynamically-derived {\it total} {\em M/L} scale is compared with the
lensing results, finding consistency (i.e., they have roughly
cross-checked the dynamical mass scale --- e.g., Cappellari et al.'s
scale --- with the lensing scale).  A more careful, explicit test of
the color-derived stellar mass scale with the lensing mass scale is
clearly warranted.

\section{Conclusions}

\begin{figure}[th]
\ \ \mbox{\includegraphics[width=9.cm,height=8cm,viewport=-14 14 514 522]{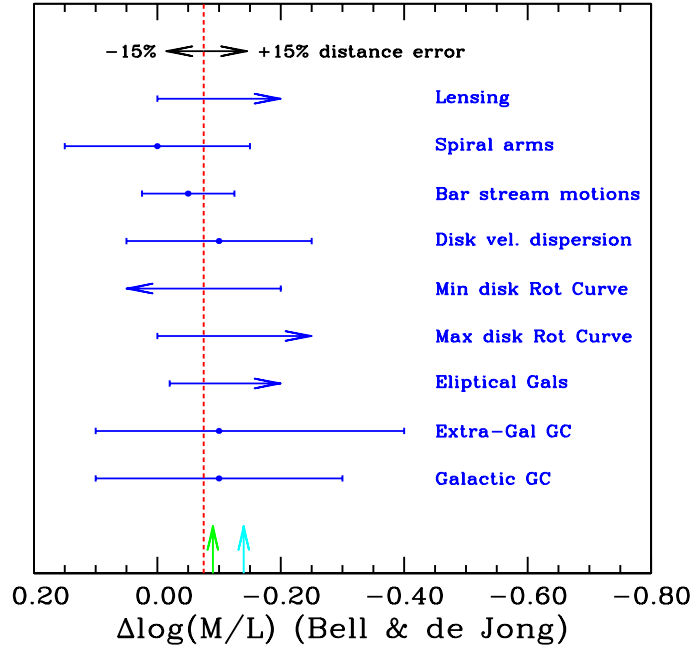}}
\caption{Overview of offsets between dynamical and stellar population
  mass estimates relative to the normalization of Bell \& de
  Jong. Ranges are indicated by horizontal errorbars, upper and lower
  limits are indicated with horizontal arrows. The vertical dotted
  line indicates the our best estimate for the offset given all
  constraints. The uncertainty in this offset caused by 15\% distance
  errors are indicated at the top. The vertical arrows indicate
  offsets expected for a Kroupa (2001) IMF (left, green) and a Kennicutt
  (1983) IMF (right, cyan).
\label{MLsummary}
}
\end{figure}

In Fig.\,\ref{MLsummary} we give an overview of all constraints
derived in the previous section on stellar population {\em M/L} values
relative to an IMF normalization of Bell \& de Jong (2001). The
strongest constraints are currently provided by the Weiner et
al.~(2001, 2004) constraints from bar streaming motions. However, this
constraint is derived from only two galaxies, and is therefore very
susceptible to for instance errors in the distances to the
galaxies. We have indicated in Fig.\,\ref{MLsummary} the effect of
15\% distance errors, which accounts for 0.075 dex offset in the {\em M/L}
IMF normalization. Another source of uncertainty lies in the stellar
population modeling. The Padova isochrone tracks (Girardi et al.~2002) are used in
many of the popular models (e.g., Charlot \& Bruzual 2003; Fioc \&
Rocca-Volmerange 1997; Vazdekis 1999) and these models all give
very similar results in terms of the color--{\em M/L} relation. However, a
class of models based on the fuel consumption theorem that treats the
late stages of stellar evolution differently are giving somewhat
different results, especially in the infrared where AGB and RGB stars
dominate (e.g., Maraston 2005). These models have indeed a slightly
steeper slope in the color--{\em M/L} relation, especially in the
near-infrared, but surprisingly the normalizations of the different
sets of models are very consistent in the intermediate color range
where most of the normalization constraints are derived.
Combining all constraints, we argue that the most
likely normalization is about 0.05--0.1 dex lower than the Bell \& de
Jong (2001) normalization, i.e.\ a Salpeter x=1.35 IMF between 0.1 and
125 $M_\odot$ reduced in mass by a factor 0.6 (-0.22 dex).

This normalization is consistent with for instance the Kroupa (2001)
or Chabrier (2003) IMFs (offset by about -0.25 dex with respect to 
Salpeter IMF) and a 
Kennicutt (1983) IMF (-0.3 dex). It leaves however not much margin
for other unseen but known mass components in the dynamical
comparisons. Stellar and substellar objects below our 0.1 $M_\odot$\ mass
limit could contribute up to 10\% (0.04 dex) of the IMF mass. A
significant molecular gas component in the inner part of spiral
galaxies could push down the rotation curve constraint. However, it is
satisfying to see that the current Galactic IMF estimates give stellar
mass estimates in a wide range of objects that are fully consistent
with their dynamical masses.

%\begin{acknowledgments}
%Colleagues...
%\end{acknowledgments}

\end{document}